\documentclass[aps,prl,amsmath,amssymb,twocolumn,letterpaper,%
         footinbib,bibnotes,showpacs,reprint,nobalancelastpage,raggedbottom,%
         citeautoscript,floatfix]{revtex4-1}
\usepackage[usenames,dvipsnames,table]{xcolor}
\usepackage[%
  breaklinks,             %
  bookmarks = false, %
  pdfpagemode   = UseNone,%
  pdfstartview  = FitH,     %
  pdfstartpage  = 1,      
  colorlinks    = true,   %
]{hyperref}
\hypersetup{
    linkcolor=RubineRed,          
    citecolor=ForestGreen,        
    filecolor=Mulberry,      
    urlcolor=RoyalBlue           
}
\usepackage{amssymb}
\usepackage{graphicx}
\usepackage{float}
\usepackage{enumerate}
\usepackage{microtype}
\usepackage{xspace}



\newcommand{\3}{$_3$}
\newcommand{\6}{$_6$}

\newcommand{\ag}{$A_g$\,}

\newcommand{\neel}{N\'{e}el}

\begin{document}


\title{Nonlinear Phononic Control and Emergent Magnetism in Mott Insulating Titanates}

\author{Mingqiang Gu}
  \affiliation{Department of Materials Science and Engineering, Northwestern University, Evanston, Illinois 60208, USA}

\author{James M.\ Rondinelli}
\email{jrondinelli@northwestern.edu}
  \affiliation{Department of Materials Science and Engineering, Northwestern University, Evanston, Illinois 60208, USA}

\begin{abstract}
Optical control of structure-driven magnetic order offers a platform for magneto-optical terahertz devices. We control the magnetic phases of $d^1$ Mott insulating titanates using nonlinear phononics to transiently perturb the atomic structure based on density functional theory (DFT) simulations and solutions to a lattice  Hamiltonian including nonlinear multi-mode interactions. We show that magnetism is tuned by indirect excitation of a Raman-active phonon mode, which affects the amplitude of the TiO$_6$  octahedral rotations that  couple to static Ti--O Jahn-Teller distortions, through infrared-active phonon modes of LaTiO$_3$ and YTiO$_3$. The mode excitation reduces the rotational angle, driving a magnetic phase transition from ferromagnetic (FM) to $A$-type antiferromagnetic (AFM), and finally a $G$-type AFM state. This novel $A$-AFM state arises from a change in the exchange interactions and is absent in the bulk equilibrium phase diagram, but it emerges as a dynamically accessible optically induced state under multi-mode excitations. Our work shows nonlinear phononic coupling is able to stabilize phases inaccessible to static chemical pressure or epitaxial strain. 

\end{abstract}

\maketitle

Recent advances in laser sciences enable light pulses to selectively pump phonon modes in crystals as a means 
to manipulate the transient atomic structure and structure-derived properties of materials. 
Ultrafast phononic control provides an alternative route beyond static methods, i.e.\ chemical pressure and thin film 
strain, to access nonequilibrium states \cite{Forst_2015Accounts}. 
By exciting an infrared-active (IR) mode so  intensively that it induces a displacive force to a Raman mode 
through a nonlinear phononic (NLP) interaction, the charge-ordering state in manganites has been melted \cite{Rini_Nature2007}, the superconducting temperature in cuprate and fullerene systems has been increased  \cite{Mitrano_Nature2016, Mankovsky_Nature2014}, and the direction of the electric polarization  in a ferroelectric has been flipped \cite{Mankowsky_Arxiv2017}---all processes on the picosecond time scale. 
Recent work has also shown that a two-mode excitation approach may be used to control the direction of a targeted distortion using polarized light \cite{Juraschek_PRL}, broadening the prospect of ultrafast structural manipulation.

In addition to well-defined electronic and dielectric states, the magnetic order in complex transition 
metal oxides are exceedingly sensitive to subtle changes in atomic structure. Indeed, control over  atomic structure through strain engineering or compositional changes can produce new ferroic states in manganites \cite{Satadeep_PRL2009, Rao_2004DeltaTran}, ruthenates \cite{Gu_PRL2012}, and titanates \cite{Lee_Nature2010}; however, 
structure-induced magnetic phase transitions in complex ternary oxides 
remain to be designed using NLP control, i.e.\ magnetophononics \cite{Wall_PRL2009,2017arXiv170703216F}, to enable dynamical multiferroism \cite{Juraschek_PRM2017, Juraschek_Sci2017}.  The main concept is to use light to exploit the structure-magnetic state dependencies originating in microscopic metal-oxygen-metal bond angles and metal-oxygen bond lengths \cite{Rondinelli/May/Freeland:2012}. In this sense, effective magnetic fields can be elicited through pure phononic excitation \cite{Nova_NatPhys2017}.

In this work, we demonstrate a  protocol to manipulate the magnetic order in ternary oxides through ultrafast dynamical structure control.
The prerequisite for such control relies on selection of an equilibrium material with its magnetic phase stability dependent upon a cooperative atomic displacive mode that resembles a natural  Raman-active mode of the crystal, e.g.\ rotations of octahedra. 
The sensitivity of the phase stability is assessed through local pertubations to the atomic structure, and then the  Raman mode that most resembles the local atomic distortion is targeted for mode-selective pumping through the  NLP interactions.
We demonstrate this process for the Mott insulating titanates and show that the NLP interactions provide access to both ferromagnetism (FM) and antiferromagnetism (AFM), including a layered $A$-type AFM absent from the equilibrium titanate phase diagram.
Although this phase has been theoretically predicted to be accessible with strain engineering \cite{Xin_JAP2013}, it remains to be observed in experiment and may be easier to achieve dynamically as proposed herein.
Finally, we show the critical laser intensity to drive the transitions can be tuned with thin film epitaxy, motivating nonlinear magnetophononics experiments on thin film titanates.

Ultrafast phononic structure control arises from the nonlinear terms in the Hamiltonian expressed as a function of the amplitudes of  Raman-active ($A_g$) and IR-active phonon ($B_u$) modes \cite{PhysRevB.89.220301}:
\begin{eqnarray}\label{eq:nonlinear}
E&=&\frac{1}{2}\nu_R^2Q_R^2 + \frac{1}{2}\nu_{IR}^2Q_{IR}^2 \\  
&+&  \frac{1}{3}a_3Q_{R}^3+ \frac{1}{4}b_4Q_{IR}^4  +  gQ_{R}Q_{IR}^2\,, \nonumber
\end{eqnarray}
where $\nu_{R,IR}$ and $Q_{R,IR}$ are the frequencies and amplitudes of the Raman and IR modes, respectively. The vibrational center of the Raman mode, which strongly affects macroscopic properties in complex oxides owing to changes in bond angles and lengths, is displaced by the IR mode through the nonlinear term $\sim Q_RQ_{IR}^2$.  By coherently pumping the IR mode one is able to manipulate the material properties on the   $10\sim100$ picosecond timescale.  

\begin{figure*}
\centering
\includegraphics[width=0.8\textwidth]{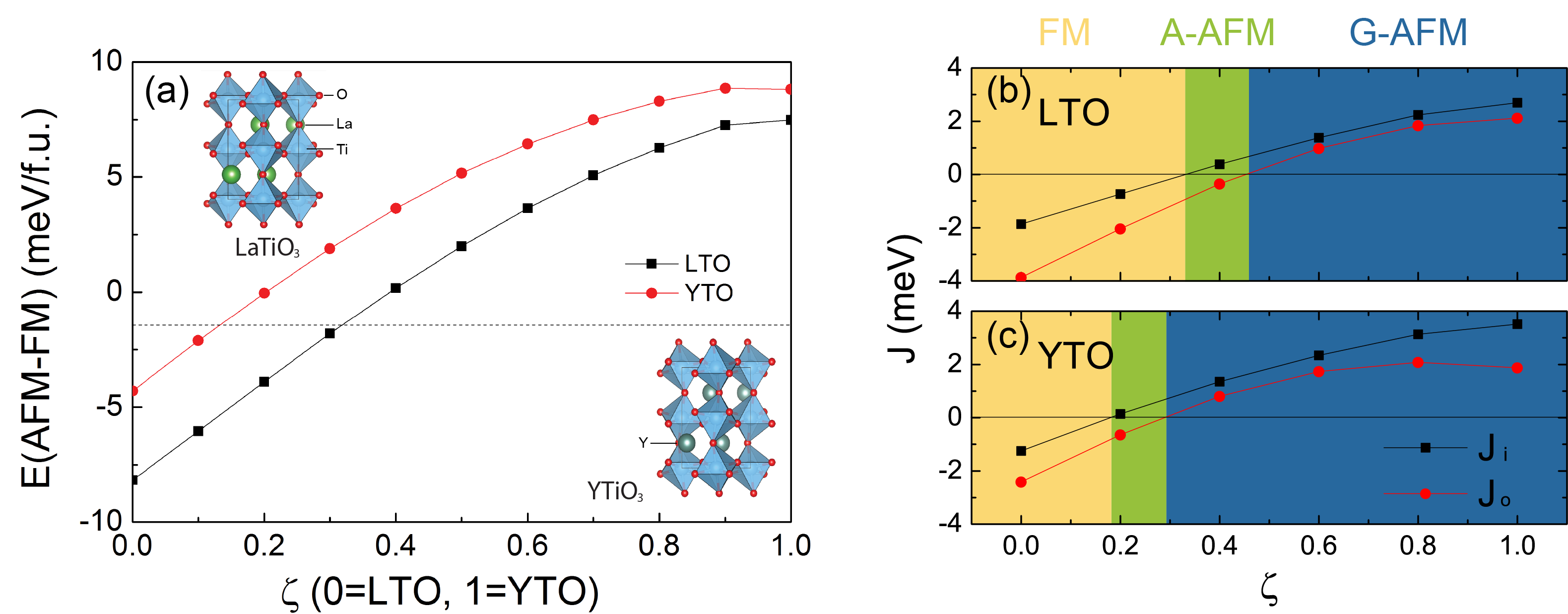}
  \caption{\label{fig:fig1} (Color online) (a) Energy difference between FM and $G$-AFM phase as a function of $\zeta$. (b,c) The exchange coupling constants between nearest  in-plane ($J_i$) and out-of-plane ($J_o$) Ti atoms as functions of $\zeta$ in LTO and YTO. Different shaded regions in (b) and (c) denote the evolution in stable magnetic phase.  The insets in (a) show the structure of LTO and YTO.}
\end{figure*}

Titanates with one electron occupying the $d$ orbital ($d^1$) are a model family to explore complex structure-dependent electronic and magnetic properties \cite{Pavarini_NJP2005}. Both LaTiO\3 and YTiO\3 are Mott insulators  \cite{Okimoto_PRB1995}, however, they exhibit different magnetic \cite{Goral_JMMM1983, Greedan_JLCM1985} and orbital \cite{Pavarini_NJP2005, Keimer_PRL2000} ordered states. 
The ground state of LaTiO\3 (LTO) is a $G$-type antiferromagnet (AFM) with a \neel\, temperature $T_N=150$ K  while YTiO\3 (YTO) is an unusual ferromagnetic (FM) insulator with a Curie temperature $T_C=30$ K. 
The appearance of both FM and AFM spin configurations indicates the $d^1$ system is not a simple Mott insulator. Prior theoretical studies have reproduced the correct magnetic ground states \cite{MOCHIZUKI1, *MOCHIZUKI2, *MOCHIZUKI3,*PhysRevB.84.195127} and identified their stability to depend on lifting of the $3d$-$t_{2g}$ orbital degeneracy, which is linked to the amplitude of the in-phase and out-of-phase TiO$_6$ octahedral rotations \cite{Pavarini_NJP2005}. 
Such octahedral rotations are enhanced in YTO owing to the small Y cation size. The magnetic ordering has been  shown theoretically to be tunable with strain \cite{Xin_JAP2013}; however, there is no experimental demonstration of the change in magnetism. Importantly, the $A$-type AFM state is unreported in any known $d^1$ titanate with trivalent $A$ site cations.

We first analyzed the relationship between the orthorhombic structure parameters and the magnetic order using first-principles calculations based on density functional theory (DFT) \cite{Supplemental_Note}. 
Both LTO and YTO exhibit $Pbnm$ symmetry whereby neighboring TiO\6 octahedra rotate about the $c$ axis in-phase ($a^0a^0c^+$) and tilt in an out-of-phase sense about the pseudocubic [110] direction ($a^-a^-c^0$ tilt pattern). 
The tilting ($\phi$) and the rotational ($\theta$) angles are defined as $\phi=[180^\circ-\angle(\mathrm{Ti}-\mathrm{O}-\mathrm{Ti})]/2$, and $\theta=[90^\circ-\angle(\mathrm{OOO})]/2$, where the reported Ti--O--Ti and interoctahedral O--O--O (denoted OOO) angles  follow the convention introduced in Ref.\ \onlinecite{Zayak_PRB2006}. 
Our calculated structure parameters for both titanates are in good agreement with the experimental data \cite{Supplemental_Note}.

Because Y exhibits a smaller cation radius than La, the unit cell volume for YTO is 8\% smaller than that of LTO.  One consequence of this is that the Goldschmidt tolarance factor $\tau=(r_\mathrm{A}+r_\mathrm{O})/[\sqrt{2}(r_\mathrm{B}+r_\mathrm{O})]$ is smaller for YTO than LTO (and both are smaller than 1). Thus, the rotation and tilt angles in YTO are larger than those in LTO. 
The change in magnitude of the octahedral rotation amplitudes in the equilibrium structures affects 
the effective exchange coupling between neighboring Ti ions through superexchange interactions.
If the total exchange coupling for nearest neighbors is written as $J=J_0+J_\mathrm{SX}$, where $J_0>0$ is the direct exchange coupling between two ions and $J_\mathrm{SX}$ is the superexchange between the two Ti ions bridged by an oxide ion, then according to the Goodenough-Kanamori-Anderson rules, $J_\mathrm{SX}$ should be negative for the $d^1$ system. This interaction then stabilizes an AFM spin configuration. Furthermore, the magnitude of superexchange $|J_\mathrm{SX}|\propto (1-\cos{\angle(\mathrm{Ti}-\mathrm{O}-\mathrm{Ti'})})$, decreases as the $\angle(\mathrm{Ti}-\mathrm{O}-\mathrm{Ti'})$ deviates from 180$^\circ$. Therefore when the rotational distortions increase, the spin system will favor a FM configuration, and vice versa. These aforementioned exchange dependencies  on the octahedral rotation angles is at the origin of the equilibrium magnetic phases in LTO and YTO.

To validate the structural origin of the magnetic configurations, we performed DFT calculations on hypothetical structures that follow an adiabatic trajectory  connecting the LTO and YTO structures. Along this trajectory both chemical compositions are used to compute the difference in total energy between the known FM and $G$-AFM spin orders (\autoref{fig:fig1}a).
Formally, we define $\zeta$ as the independent structural parameter in the trajectory, such that $\zeta=0$ ($\zeta=1$) denotes the equilibrium LTO (YTO) structure. Structures between $\zeta=0$ and 1 are obtained as a linear interpolation between the two end members, i.e. $x_\zeta(i)=x_0(i)+\zeta dx(i)$, in which $x(i)$ is the fractional coordinate for atom $i$ and $dx(i)=x_1(i)-x_0(i)$. 
We find for both compounds that independent of the La or Y chemistry, if the titanate exhibits the LTO crystal structure ($\zeta=0$) than the $G$-AFM state is always favored. 
As $\zeta$ increases away from zero in the LTO structure towards that of  YTO, a magnetic transition occurs at $\zeta\sim0.15$ for Y and $\zeta\sim0.3$ for La. At these values the FM state is energetically favored. 

Microscopically, these changes in magnetic states are due to changes in the effective exchange coupling $J$.
To identify the different contributions from the rotation and tilt distortions, which are related to the equatorial or apical oxide anions connecting the two Ti cations, respectively, we computed the in-plane ($J_i$) and out-of-plane ($J_o$) exchange constants between nearest Ti sites (\autoref{fig:fig1}b). 
It is clear that the signs for both  $J_i$ and $J_o$ change from negative (AFM coupling) to positive (FM coupling) as the structure evolves from that of LTO to YTO. The critical $\zeta$ for the in-plane coupling is about 0.2 while that for the out-of-plane is $\sim0.3$ in YTO. These critical values shift to higher $\zeta$ for LTO and these findings are consistent with the phase diagram given in \autoref{fig:fig1}. 
Remarkably, our analysis shows that there should be an additional $A$-type AFM phase bridging the FM and $G$-AFM phases in the region of $0.2<\zeta<0.3$. The total energy for the $A$-AFM configuration at $\zeta=0.25$ is about 0.5 meV more stable than that of the FM phase.

The critical rotation and tilt angles required to achieve the transition are smaller than the average structure ($\zeta=0.5$). In LTO these angles are $\theta=11.4 ^\circ$, $\phi=17.5^\circ$ (FM to $A$-AFM) and $\theta=11.7 ^\circ$, $\phi=18.3^\circ$ ($A$-AFM to $G$-AFM), respectively, which are close to those for YTO: $\theta=11.0 ^\circ$, $\phi=16.6^\circ$ (FM to $A$-AFM) and $\theta=11.3 ^\circ$, $\phi=17.3^\circ$ ($A$-AFM to $G$-AFM), respectively.
Therefore, we propose that control of the magnetic state in the $d^1$ titanates should be possible through changes in the TiO$_6$ octahedral rotation and tilt angles. 
Since these rotational modes transform as the fully symmetric irreducible representation of the point group $mmm$, the next task is to identify the  Raman mode that is optimally suited for optical pumping via the NLP interaction.

Using YTO as an example, we identify the Raman mode that most effectively modulates the octahedral rotation amplitude by computing the lattice dynamical properties of YTO and checking the mode similarity among seven Raman-active \ag modes to pure octahedral rotation, tilt and Jahn-Teller distortions \cite{Supplemental_Note}.
The \ag mode (index no.\ 25, $\nu=298$ cm$^{-1}$) exhibits the largest composition of octahedral rotation.
Furthermore, the energy difference $\Delta E(\mathrm{AFM}-\mathrm{FM})$ drops rapidly when this mode is excited (see Fig.\ S1 of Ref.\ \onlinecite{Supplemental_Note}), which confirms our selection of this mode to tune the magnetic order.

Note that the phase diagram obtained by exciting the \ag\,(25) phonon mode shown in Fig.\ S3 of Ref.\ \onlinecite{Supplemental_Note} differs from that obtained
with respect to the structural parameter $\zeta$ (\autoref{fig:fig1}), because the Raman mode is not identical to $\zeta$. 
The contributions of pure octahedral rotation, tilt and Jahn-Teller distortions differ between the \ag\,(25) phonon and $\zeta$ parameter; there is more rotation present in  \ag\,(25). 
%
Although the energies for the FM, $G$-AFM, and $A$-AFM configurations are close in YTO, the FM state is always energetically favored over the $A$-AFM order; in addition, the FM to $G$-AFM transition occurs at $Q=1.65 \mathrm{\AA \sqrt{amu}}$ (Fig.\ S3), which indicates that the $A$-AFM state remains hidden by the excitation of the $A_g$ mode alone.
Interestingly, the \ag mode with $\nu=512$ cm$^{-1}$ (index no.\ 49) is almost a pure Jahn-Teller mode \cite{Supplemental_Note}.  Although this mode is not efficient in reducing the energy difference between the FM and $G$-AFM, this Q$_2$ Jahn-Teller mode is the same mode that stabilizes the $A$-AFM magnetic order in LaMnO\3 \cite{LMO}, and therefore may influence the stability of the $A$-AFM phase relative to the FM and $G$-AFM order if it is also excited during the NLP process.

\begin{figure}
\centering
\includegraphics[width=0.47\textwidth]{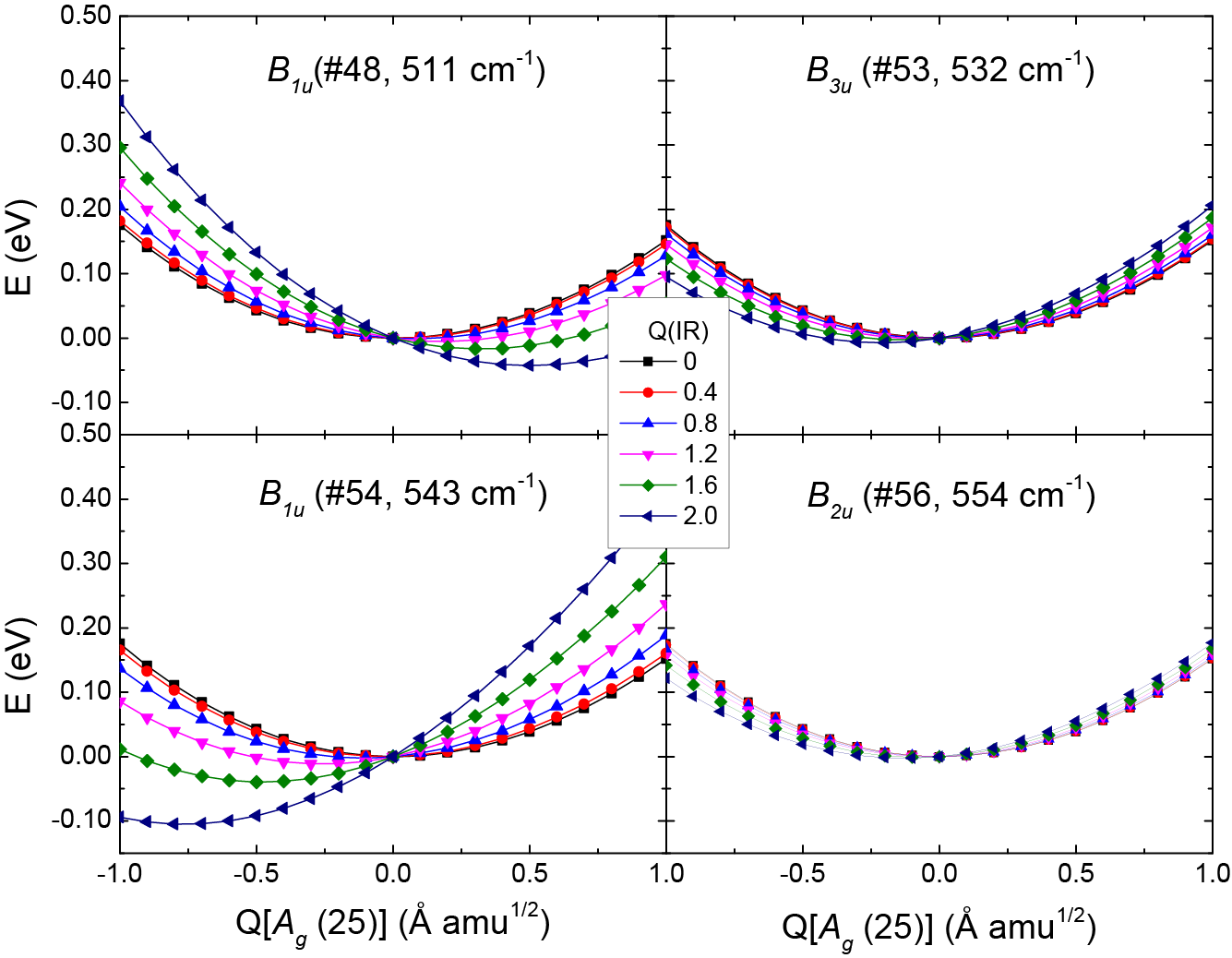}
  \caption{\label{fig:figure3} Energy profiles for the nonlinear coupling between IR-active modes and the \ag (25) Raman mode in YTiO$_3$.}
\end{figure}

Next, we identify which IR mode will most effectively couple to the $A_g$(25) Raman mode through a NLP interaction. According to the theory of ionic Raman scattering, the amplitude of the Raman mode is largest  when the frequency difference between the driven Raman mode and the pumped IR mode is maximized  \cite{Mankowsky_PRB.91.094308}. Therefore the coherently pumped IR mode should be a high frequency mode albeit accessible to current laser THz-laser sources. 

We now examine the nonlinear coupling interactions between the Raman \ag(25) mode and five $B_u$ IR-active modes with the highest frequencies. (The frequencies and phonon characters for the five IR modes are given in Ref.\ \onlinecite{Supplemental_Note}). 
By fitting the energy surfaces $E(Q_\mathrm{R}, Q_\mathrm{IR})$  to \autoref{eq:nonlinear}, we obtained the coupling coefficients $g$, in Table S2 of Ref.\ \onlinecite{Supplemental_Note}. We find four IR modes with a nonlinear coupling coefficient near or larger than 0.01 eV/$(\mathrm{\AA}\sqrt\mathrm{amu})^3$. The energy profiles for each of these  modes with the \ag\,(25) mode are shown in \autoref{fig:figure3}. 
When the $B_{1u}$ mode (index no.\ 48, $\nu=511$ cm$^{-1}$) is pumped to an amplitude of 2 \AA $\sqrt{\text{amu}}$, the \ag(25) mode finds its energy minimum at a nonequilibrium value $Q(A_g) \sim 0.5$ \AA $\sqrt{\text{amu}}$. With the excitation of the $B_{3u}$ mode (index no.\ 53, $\nu=532$ cm$^{-1}$), the energy minimum of the \ag(25) mode is displaced by $\sim -0.2$ \AA $\sqrt{\text{amu}}$. A similar strength excitation of another $B_{1u}$ mode (index no.\ 54, $\nu=543$ cm$^{-1}$) displaces the energy minimum by $\sim -0.8$ \AA $\sqrt{\text{amu}}$. 
%
For the weakly coupled $B_{2u}$ mode (index no.\ 56, $\nu=554$ cm$^{-1}$), 2 \AA $\sqrt{\text{amu}}$ excitation only shifts the energy minimum to $approximately -0.1$ \AA $\sqrt{\text{amu}}$.

Among these IR modes, the nonlinear coupling coefficient for the $B_{1u}$\,(48) and the $B_{1u}$\,(54) are comparable. However, only the $B_{1u}$\,(48) mode shifts the energy minimum of the \ag(25) mode towards a larger amplitude, which is required to \emph{reduce} the octahedral rotation angles and drive the magnetic transition. Therefore, this IR mode is selected to drive changes in the Raman mode through the NLP interaction. 
Note that the coefficient $g[A_g(25),B_{1u}(48)] \sim -0.039$ eV/$(\mathrm{\AA}\sqrt\mathrm{amu})^3$, which is three times larger in magnitude compared to the nonlinear phononic coupling strength in LaTiO\3  \cite{Gu_PRB2017}. 
This value is approximately one order of magnitude smaller than that reported in YBa$_2$Cu$_3$O$_7$ \cite{Fechner_PRB2016} and about half of that in PrMnO\3 \cite{PhysRevB.89.220301}.

\begin{figure}
\centering
\includegraphics[width=0.4\textwidth]{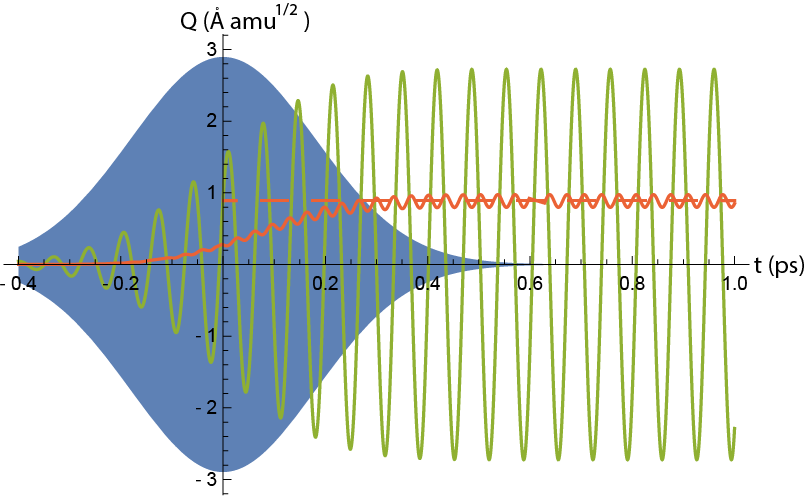}
  \caption{\label{fig:figure4} Time evolution of the \ag (25) mode (red) and the $B_{1u}$\,(48) mode (green). The laser pulse envelope is schematically shown as the blue filled-curve. The dashed line denotes the displacive vibrational center of the \ag (25) mode after pumped by a laser pulse with $\omega=14.5$ Thz and $I=18$ MV/cm (without considering damping). }
\end{figure}

We compute the time evolution of the phonon modes during this process by solving the equations of motion expressed in Eq.\ S1 of Ref.\ \onlinecite{Supplemental_Note}. For efficient mode pumping, the laser frequency should be in a resonant condition with the IR mode. Note that it is important to consider the red-shift of the IR mode in the nonlinear process under high laser intensity \cite{Fechner_PRB2016}; therefore, we propose the laser frequency of $\omega= 14.5$ THz. 
With a laser intensity of 18 V/cm, the time evolution of the \ag(25) and $B_{1u}$(48) modes are plotted in \autoref{fig:figure4}. The maximum displacement of the \ag(25) mode is $\sim 0.9$\,\AA $\sqrt{\text{amu}}$ corresponding to a rotational angle reduction of $\sim 1.6 ^\circ$; however, the excited mode amplitude is insufficient to drive the magnetic transition. In order to dynamically achieve  it, one should either increase the laser intensity or reduce the critical rotation angle through additional means. We discuss both options below.

\begin{figure}
\centering
\includegraphics[width=0.4\textwidth]{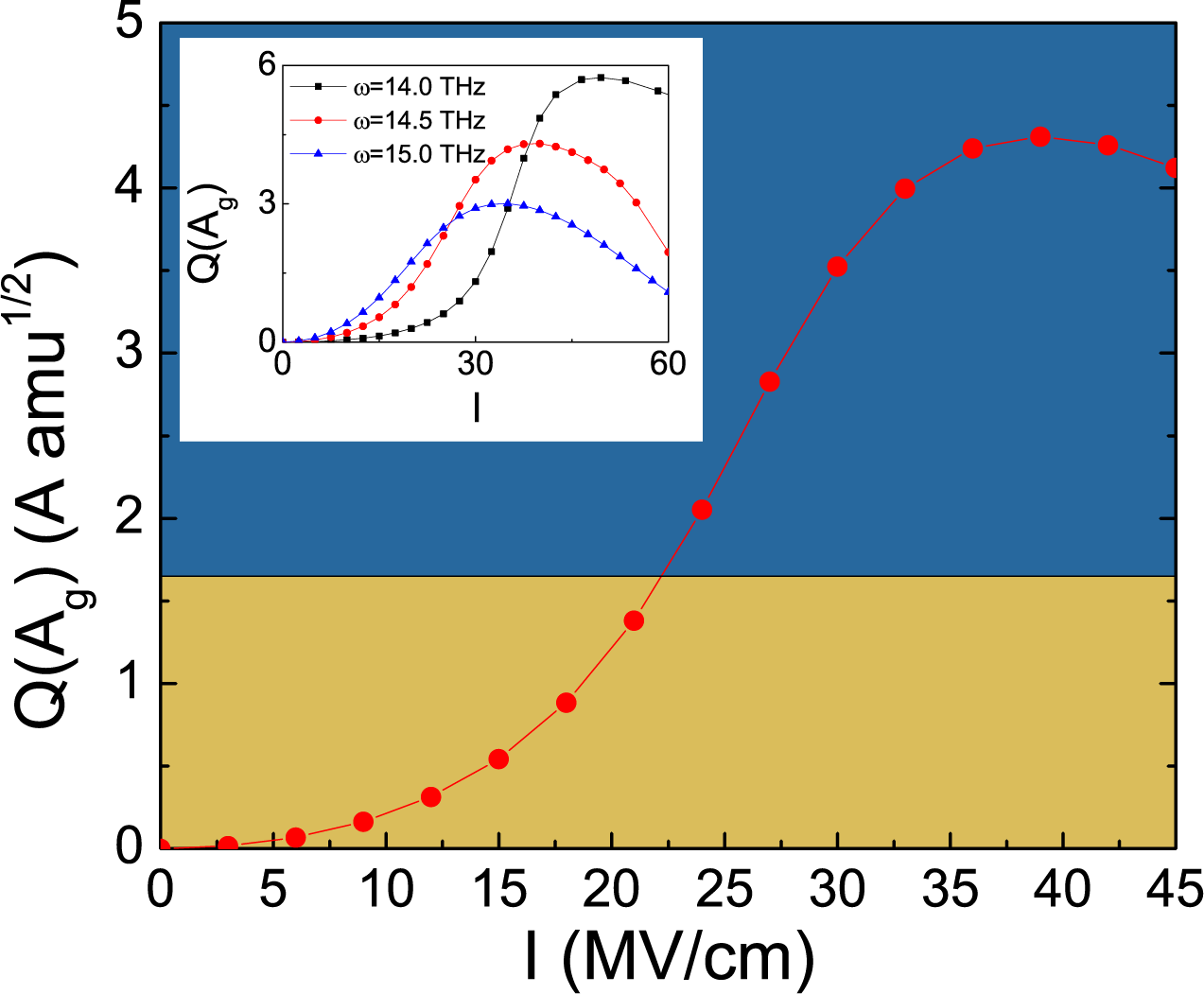}
  \caption{\label{fig:figure5} Mode amplitude $Q(A_g)$ and the change in rotation angle $\Delta\phi$ as a function laser intensity ($\omega=14.5$\,THz). Different colors denote different magnetic order as in \autoref{fig:fig1}. The inset shows the same quantity for different pump frequencies.}
\end{figure}

First we assess the relationship among the laser parameters (frequency and intensity) and the resulting driven mode amplitude in more detail to 
understand how strongly the mode should be driven to achieve the dynamical magnetic transition.
\autoref{fig:figure5} shows the stationary nonequilibrium displacements (damping is neglected in this discussion) of the \ag(25) mode as a function of the laser intensity for different pump frequencies. We find that at low laser intensity, a higher frequency pump induces a larger displacive amplitude of the Raman mode whereas lower pump frequencies lead to larger maximum displacements \autoref{fig:figure5}. The reason for this counterintuitive result relies on the frequency red-shift of the IR mode during the NLP process, which is discussed in Ref.\ \onlinecite{Supplemental_Note}. We find that the estimated critical laser intensity from \autoref{fig:figure5} with $\omega=14.5$ THz is $\sim$22 MV/cm.

Although one can increase the laser intensity to acquire a larger displacive Raman mode, the laser damage threshold of the material limits the largest laser intensity that can be used in the pump-probe experiment \cite{Gu_2016SciRep}. Also, sample heating due to an intense laser pulse is problematic for low-temperature measurements---here the experiments should be performed below the bulk magnetic ordering temperatures of the titanates. A more practical way to experimentally observe the dynamical phonon-induced magnetic phase transition is to reduce the critical rotation angle required for the phase transition

Here we propose to impose static epitaxial strain on a YTO thin film grown on a (110)-oriented substrate. This mechanical constraint is based on the fact that both the rotation and tilt angle should be reduced to bring the material close to the phase transition point. If the sample is strained in the conventional [001]  orientation, the changes in the rotation and the tilt angles have an opposite behavior and effectively cancel: For compressive (tensile) strains, the rotation angle increases (decreases) while the tilt angle decreases (increases). 
If the sample is grown on a (110)-oriented substrate, and strain is applied along [111] and [1$\bar{1}$1] directions (in terms of pseudo-cubic axes), then both the rotation and tilt angles will decrease or increase together \cite{JMR_AM2011}. Indeed for YTO, we find both angles decrease almost linearly when tensile strain is applied \cite{Supplemental_Note}. The energy difference between FM and $G$-AFM phase is only 0.4 meV/f.u.\ at a tensile strain of 2\%, suggesting a substrate lattice constant of approximately 3.98\,\AA\, would be the best option.
Our calculations on the NLP coupling on [110]-oriented strained YTO show that $g[A_g(24),B_{1u}(51)] \sim -0.04$ eV/$(\mathrm{\AA}\sqrt\mathrm{amu})^3$ is nearly unchanged   \cite{Supplemental_Note}; however, the critical laser intensity for the magnetic transition is computed to be 13.1 MV/cm, reduced by 40\,\%.
These features make KTaO\3 ($a\sim3.988$\,\AA) a good candidate for which to grown YTO films on and attempt the magnetophononic experiment.

Another possible route to reduce the critical rotation angle is to consider the excitation of a Jahn-Teller (JT) type Raman mode. The JT mode is usually closely related to the orbital occupation and magnetic ordering in perovskites \cite{LMO}, and therefore, could dynamically bring the system closer to the phase transition boundary. Such a JT mode should be simultaneously excited through the nonlinear phononic coupling when symmetry allowed.
To assess this effect, we computed the energies for the FM, $G$-AFM, and $A$-AFM phases as functions of the \ag\,(25) mode amplitude in a background of the $Q[A_g (49)] = 0.2$   \AA $\sqrt{\text{amu}}$, which corresponds to the driven amplitude at the FM-to-$G$-AFM transitions (\autoref{fig:figure5}).
The frequency of this mode (512 cm$^{-1}$) is very close to the pump frequency and therefore its amplitude could be resonantly enhanced.

When the JT mode is also excited, the magnetic energy landscape changes and the $A$-AFM phase becomes accessible with $Q[A_g(25)] < -1.4$  \AA $\sqrt{\text{amu}}$. Note that now the direction of the displacive amplitude of the $A_g$(25) mode matters. To drive the FM to $A$-AFM phase transition, the $B_{1u}$(54) mode should be used (lower left panel, \autoref{fig:figure3}). The calculated pump-laser parameters to drive this phase transition are $\omega=16$\,THz and $I=16.8$\, MV/cm; however, if $Q[A_g (49)] = 1$ then the required laser parameters are $\omega=14.5$\,THz and $I=16.4$\,MV/cm. 
The FM to $G$-AFM phase transition occurs at values of 1.5\,\AA $\sqrt{\text{amu}}$ or greater, slightly reduced from the value with a single mode excitation [$Q[A_g (49)] = 0$]. The critical intensity is $I=21.6$ MV/cm.
Last, we emphasize the combination of these two approaches can be used to further reduced the required pump intensity to access
%
the hidden $A$-AFM phase arising in the NLP interaction.
%

\begin{acknowledgments}
The authors thank M.\ Fechner, R.\ Averitt, V.\ Gopalan, and D.\ Puggioni for fruitful discussions. M.G.\ and J.M.R.\ acknowledge financial support from the U.S.\ Department of Energy (DOE) under Grant No. DE-SC0012375. Calculations were performed using the Extreme Science and Engineering Discovery Environment (XSEDE), which is supported by NSF Grant No.\ ACI-1548562 and and the CARBON Cluster at Argonne National Laboratory (DOE-BES, Contract No.\ 
DE-AC02-06CH11357).
\end{acknowledgments}

\bibliography{references}

\end{document}